\documentclass[10pt,twocolumn]{article}
\usepackage[utf8]{inputenc}
\usepackage{amsmath,amssymb}
\usepackage{graphicx}
\usepackage{booktabs}
\usepackage{multirow}
\usepackage{array}
\usepackage{hyperref}
\usepackage{cite}
\usepackage{caption}
\usepackage{subcaption}
\usepackage{float}
\usepackage{setspace}
\usepackage{geometry}
\usepackage{xcolor}
\usepackage{lipsum}
\usepackage{stfloats} 
\title{Assessing the Effect of PCA-Based Dimensionality Reduction on Machine Learning Performance in Hyperspectral Optical Imaging}
\author{
    Parisa Parand\textsuperscript{1} \and
    Mahmoud Samadpour\textsuperscript{2}\textsuperscript{*} \\
    \small \textsuperscript{1}Dept. Statistics, Mathematics, and Computer Science, Allameh Tabataba'i University, Tehran, Iran \\
    \small \textsuperscript{2}Physics Department, K.N. Toosi University of Technology, Tehran, Iran \\
    \small \textsuperscript{*}samadpour@kntu.ac.ir
}
\date{}

\begin{document}

\maketitle

\begin{abstract}
Hyperspectral optical imaging provides rich spectral information for estimating continuous environmental and material parameters; however, its high dimensionality and strong feature correlation pose significant challenges for machine learning models, especially when ground-truth datasets are limited. In this study, we investigate a hyperspectral dataset composed of 150 spectral bands with soil moisture as the target variable. To address the curse of dimensionality, Principal Component Analysis (PCA) was employed as a baseline dimensionality reduction technique. The optimal number of principal components was determined to be two, retaining more than 99\% of the total variance. This selection was supported by the analysis of the covariance matrix, eigenvalue distribution, and the scree plot. Projecting the data onto the first two principal components enabled improved visualization and interpretability compared to the original high-dimensional feature space. The reduced representation also revealed a clearer separation of target values, effectively decreasing data complexity. To evaluate the impact of dimensionality reduction on predictive performance, a Random Forest regression model was trained to estimate soil moisture from the PCA-transformed data. The model achieved a coefficient of determination (\(R^2\)) of 94.7 \%, demonstrating that PCA-based feature reduction can enhance computational efficiency while preserving strong predictive capability in hyperspectral machine learning workflows.

\end{abstract}

\section{Introduction}
Optical imaging systems are increasingly capable of capturing large volumes of high-dimensional data, enabling detailed characterization of materials and environmental conditions across scientific and engineering domains. Among such technologies, hyperspectral optical imaging provides dense spectral information by recording reflectance or radiance values across dozens to hundreds of contiguous wavelength bands \cite{Camps-Valls2012}. These multidimensional datasets offer rich feature representations but pose significant computational and analytical challenges due to their size, redundancy, and strong band-to-band correlations \cite{Gewali2018}.

Machine learning (ML) methods have become essential tools for extracting meaningful information from high-dimensional optical datasets, supporting tasks such as material identification, biomedical tissue analysis, food quality monitoring, environmental assessment, geoscientific studies, precision agriculture, and industrial inspection \cite{Ali2015, Treitz1999}. While a substantial body of research has focused on hyperspectral classification, where targets represent discrete categories \cite{Bioucas-Dias2013}, comparably fewer studies have explored hyperspectral regression, in which the goal is to estimate continuous physical or chemical parameters from imaging data \cite{Bellman1961, Colini2014}.

Hyperspectral regression problems are often impacted by the curse of dimensionality \cite{Keller2018}, a phenomenon in which the sparsity of data in high-dimensional feature spaces increases the difficulty of model training and may require disproportionately large datasets for reliable generalization. Because hyperspectral bands exhibit strong correlations and redundant information, the intrinsic or virtual dimensionality of the data is frequently much lower than its nominal spectral resolution \cite{Chang2018}. Dimensionality reduction is therefore an essential step to enable efficient and robust hyperspectral machine learning pipelines.

Among the available techniques, PCA remains one of the most widely used linear dimensionality reduction methods due to its simplicity, interpretability, and computational efficiency \cite{Pearson1901}. PCA orthogonally transforms the original feature space into a smaller set of uncorrelated variables---principal components---ordered by their explained variance. By retaining only the most informative components, PCA facilitates noise reduction, improves model interpretability, and may accelerate machine learning tasks.

To demonstrate these concepts, this work investigates a hyperspectral optical imaging dataset with 150 spectral bands, in which soil moisture serves as an example of a continuous parameter of interest. After applying PCA to reduce the data dimensionality, a Random Forest regression model is used to evaluate the impact of dimensionality reduction on predictive performance. The model achieves a coefficient of determination (\(R^2\)) of 94.7\%, indicating that PCA-based dimensionality reduction can produce efficient feature representations while preserving strong predictive capability.

Overall, this study positions PCA as a baseline approach for improving machine learning performance in high-dimensional optical imaging scenarios and provides a template for integrating dimensionality reduction within broader computational imaging workflows.

\section{Research Methodology}

\subsection{Dataset Description}
The hyperspectral optical imaging dataset used in this study was introduced in \cite{Riese2018} and contains 679 data samples, each consisting of a continuous soil moisture value and 125 spectral bands. The dataset was acquired during a five-day field measurement campaign conducted in May 2017 in Karlsruhe, Germany. Measurements were performed on an undisturbed bare-soil sample collected near Waldbronn, Germany. A Cubert UHD 285 hyperspectral snapshot camera was employed to capture spatially resolved spectral information. The system records \(50 \times 50\) pixel images across 125 contiguous spectral bands ranging from 450 nm to 950 nm, with an approximate spectral resolution of 4 nm. Reference soil moisture values were obtained using a TRIME-PICO Time-Domain Reflectometry sensor. Table~\ref{tab:dataset} summarizes the dataset structure, which includes measurement soil moisture (\%), soil temperature (°C), and hyperspectral bands from 454 nm to 950 nm.

\begin{table*}[t]
\centering
\caption{Structure of dataset. Spectral bands (features) range from 454 to 950 nm, and soil moisture is the target variable.}
\label{tab:dataset}
\small
\begin{tabular}{@{}ccccccccc@{}}
\toprule
Sample & Soil Moisture (\%) & Soil Temp. (°C) & \multicolumn{6}{c}{Bands (nm)} \\
& & & 454 & 458 & 462 & \ldots & \ldots & 950 \\
\midrule
1 & 33.51 & 34.8 & 0.0821 & 0.0558 & 0.0500 & \ldots & \ldots & 0.1539 \\
2 & 33.49 & 35.2 & 0.0795 & 0.0553 & 0.0491 & \ldots & \ldots & 0.1567 \\
\ldots & \ldots & \ldots & \ldots & \ldots & \ldots & \ldots & \ldots & \ldots \\
679 & 29.75 & 39.7 & 0.0976 & 0.0654 & 0.0560 & \ldots & \ldots & 0.1659 \\
\bottomrule
\end{tabular}
\end{table*}

\subsection{Data Preparation}
The dataset was imported into Python using the \texttt{pandas.read\_csv()} function. The data were decomposed into the feature matrix, \(X\): 125 hyperspectral band values per sample (450--950 nm), and target vector, \(y\): soil moisture values (\%).

To ensure consistent reproducibility of experiments, a fixed random state (42) was set using \texttt{numpy.random.seed()}, controlling the internal random number generator used later for dataset splitting and model initialization. Before analysis, the distribution of soil moisture values was inspected using a histogram generated with the \texttt{DataFrame.hist()} method from Pandas, which internally calls \texttt{matplotlib.pyplot.hist()}.

\subsection{Dimensionality Reduction and Machine Learning Model}
To address the high dimensionality and strong inter-band correlation inherent in hyperspectral data, PCA was applied. All spectral features were standardized using \texttt{sklearn.preprocessing.StandardScaler}. PCA was implemented via \texttt{sklearn.decomposition.PCA}. The number of principal components was selected such that more than 99\% of the total variance was retained. This threshold resulted in two principal components, confirmed through:

\begin{itemize}
    \item Examination of the covariance matrix
    \item Eigenvalue distribution and cumulative variance analysis
    \item A scree plot to visualize variance decay
\end{itemize}

The resulting reduced representation enabled improved visualization and preserved the dominant structure of the data in a low-dimensional subspace. To evaluate the impact of dimensionality reduction on predictive performance, a Random Forest Regression model was trained on the PCA-transformed feature space. The model was implemented using \texttt{sklearn.ensemble.RandomForestRegressor}. Test-size was selected to 0.3, random-state was fixed to 42, number of trees was set to 100, and performance was quantified using the \(R^2\) score.

\section{Results and Discussion}
\begin{figure}[ht]
\centering
\includegraphics[width=0.9\linewidth]{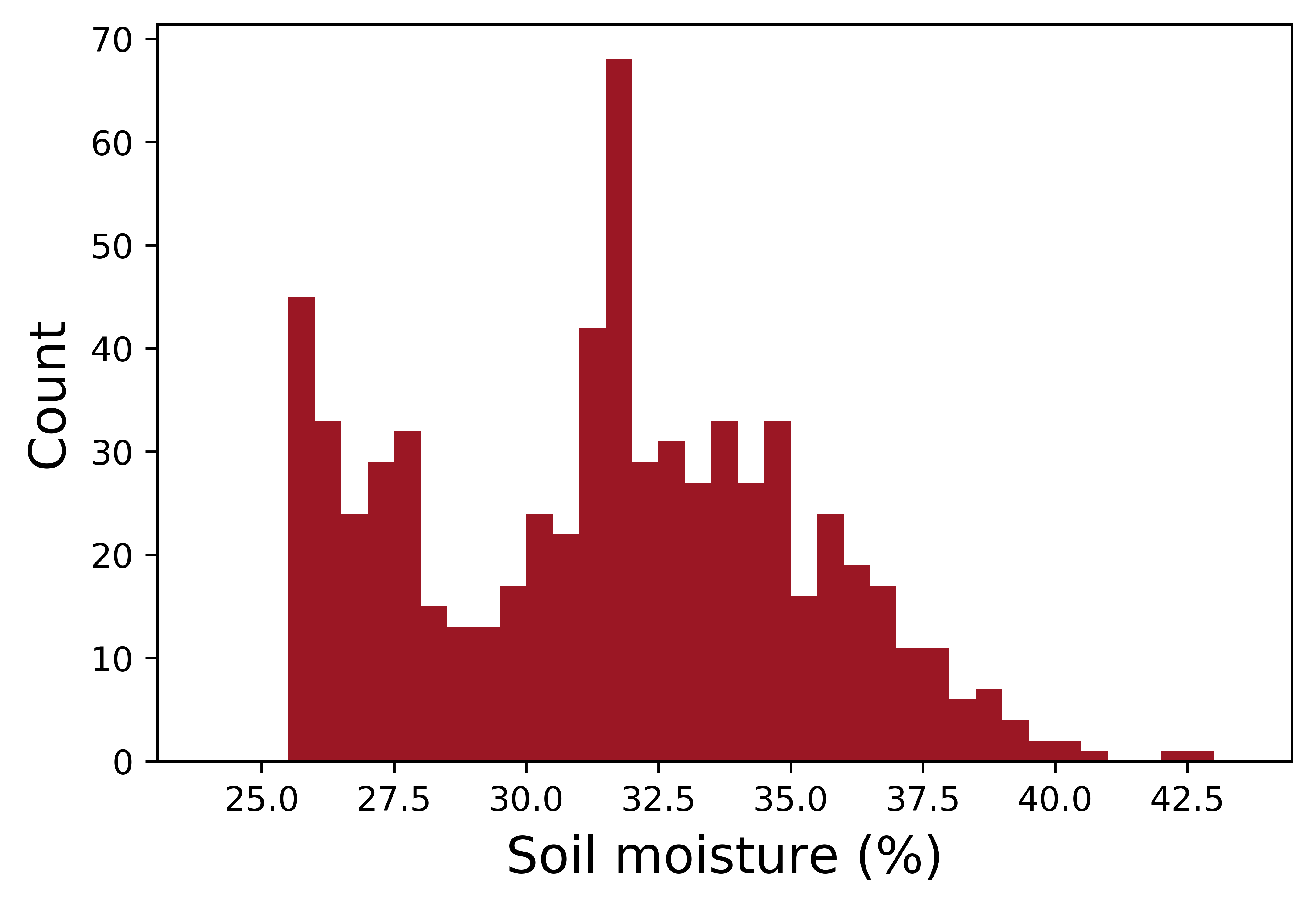}
\caption{Distribution of soil moisture in the reference dataset, showing most measurements concentrated near 32\%.}
\label{fig:dist1}
\end{figure}

Figure~\ref{fig:dist1} illustrates the distribution of soil moisture values in the reference dataset. The measurements range from approximately 25\% to 43\%, with the majority of samples centered around 32\%, indicating a moderately narrow distribution of soil moisture values across the measurement period.

\begin{figure}[ht]
\centering
\includegraphics[width=0.9\linewidth]{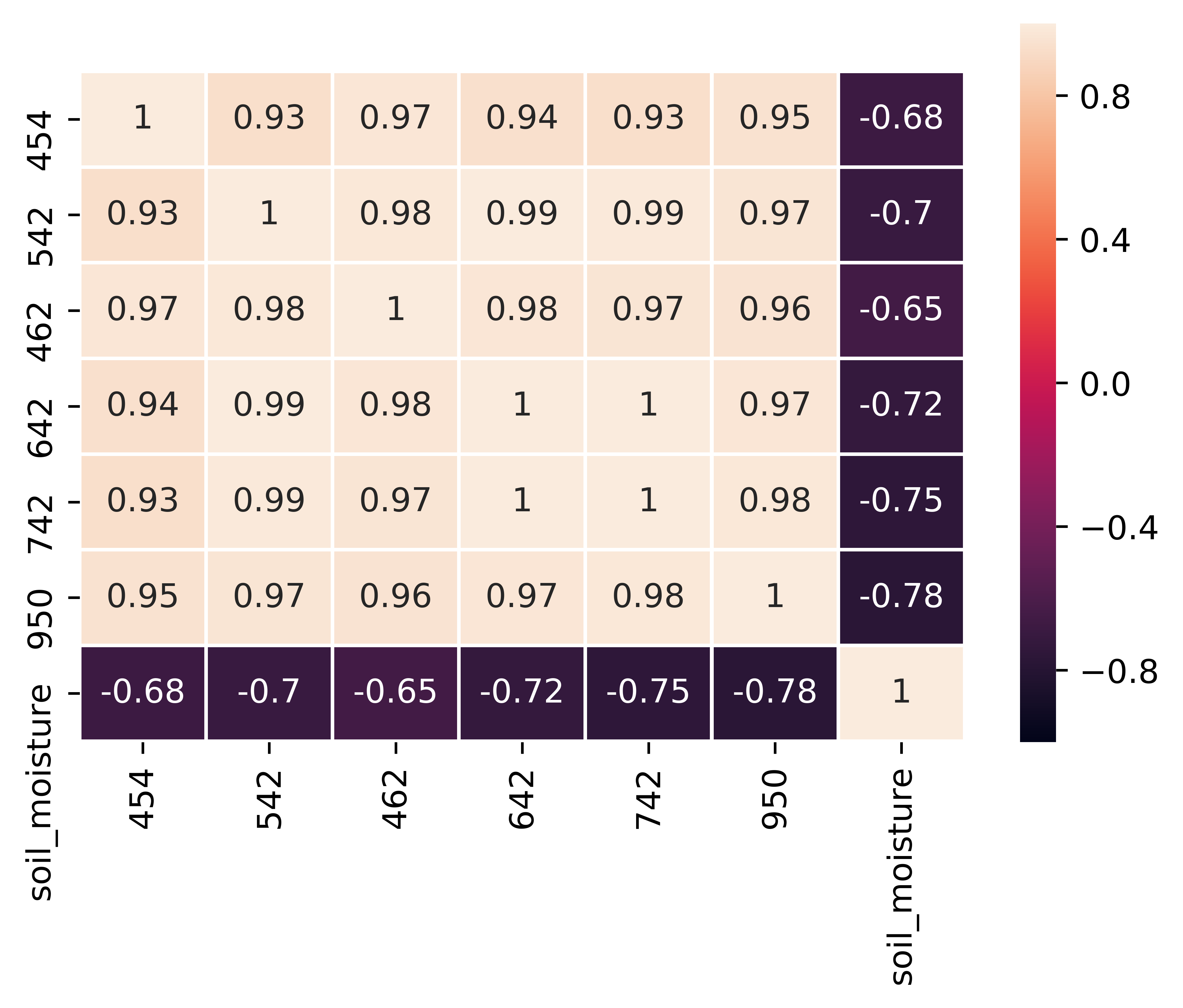}
\caption{Correlation heatmap of sample hyperspectral bands and soil moisture (correlation scale: \(-1\) to \(+1\)).}
\label{fig:heatmap}
\end{figure}

To explore the statistical relationships between spectral features and the target variable, a correlation heatmap was generated using the \textit{Seaborn} library (Figure~\ref{fig:heatmap}). For demonstration, six hyperspectral bands were selected to highlight representative correlations with soil moisture. Several bands display strong interband correlations, notably at 642 nm and 742 nm, reflecting spectral redundancy within the measured wavelength interval. Such behavior is typical in hyperspectral systems where adjacent bands often convey overlapping information. This suggests that dimensionality reduction techniques can be applied to reduce computational complexity without significant information loss, a key advantage for large-scale datasets.

\begin{figure}[ht]
\centering
\includegraphics[width=0.9\linewidth]{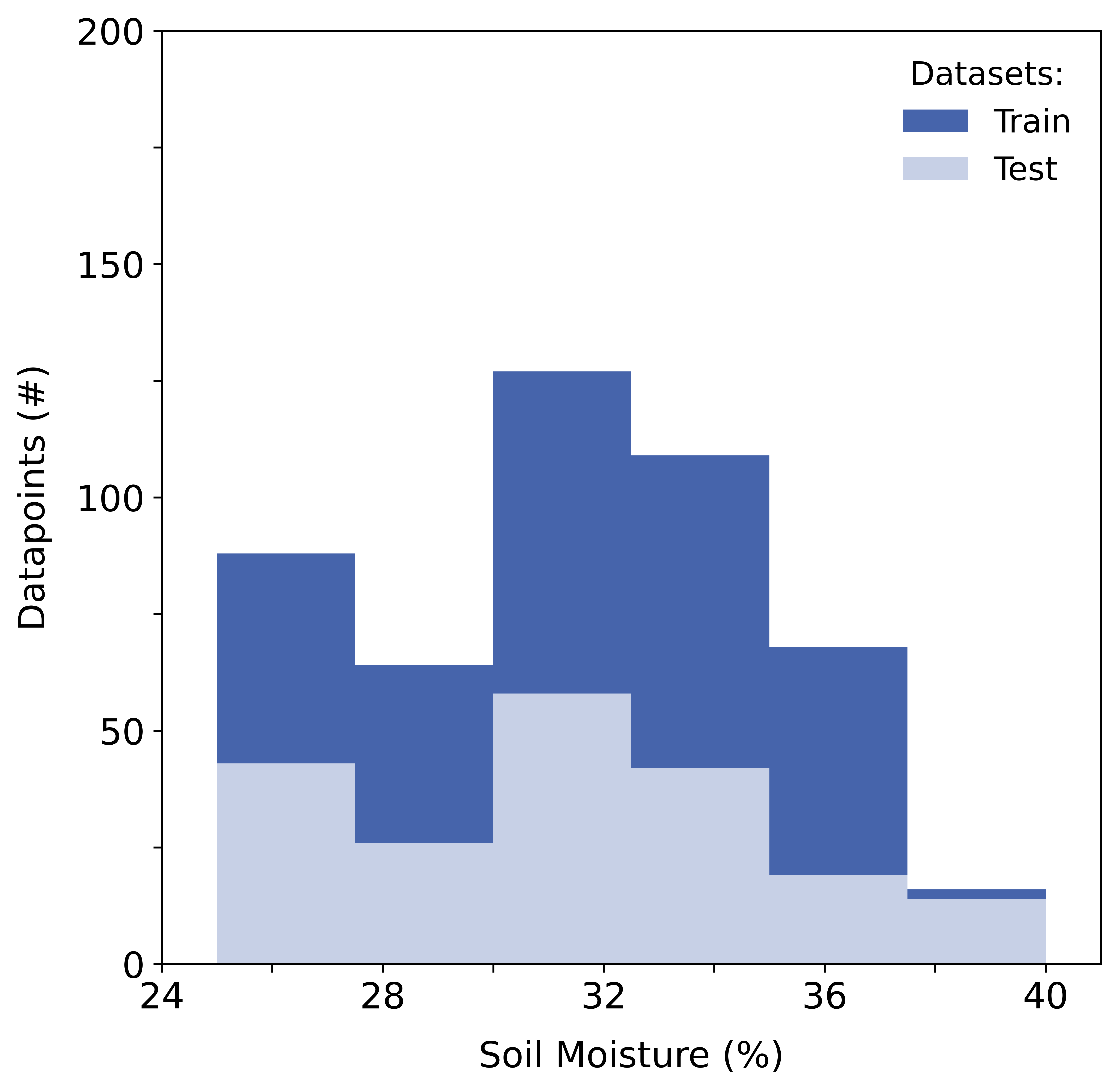}
\caption{Distribution of soil moisture in train and test samples.}
\label{fig:dist2}
\end{figure}

Before model training, the dataset was standardized using StandardScaler from scikit-learn, and subsequently partitioned into training and test subsets via \texttt{train\_test\_split()}, with a test size of 0.3 and a fixed random state of 42 to ensure reproducibility. Figure~\ref{fig:dist2} compares the distribution of soil moisture in training and test samples, demonstrating that both subsets follow closely aligned distributions, supporting a reliable basis for supervised learning.

\subsection{Dimensionality Reduction using PCA}
Since the dataset contains 125 spectral bands, it represents a high-dimensional feature space that may impede model interpretability. Therefore, PCA was performed to reduce dimensionality while preserving maximum spectral information.

The explained variance ratios for the first four principal components are presented in Table~\ref{tab:pca}. To preserve at least 99\% of the dataset's variance, retaining only PC1 and PC2 is sufficient, while higher components contribute marginally and can be excluded without meaningful information loss.

\begin{table*}[t]
\centering
\caption{Variance ratio and cumulative explained variance for first four principal components}
\label{tab:pca}
\begin{tabular}{@{}ccccc@{}}
\toprule
 & \textbf{PC-1} & \textbf{PC-2} & \textbf{PC-3} & \textbf{PC-4} \\
\midrule
\textbf{Variance ratio} & 0.9889 & 0.0064 & 0.0023 & 0.0016 \\
\textbf{Cumulative variance ratio} & 0.9889 & 0.9953 & 0.9976 & 0.9992 \\
\bottomrule
\end{tabular}
\end{table*}

\begin{figure}[ht]
\centering
\includegraphics[width=0.9\linewidth]{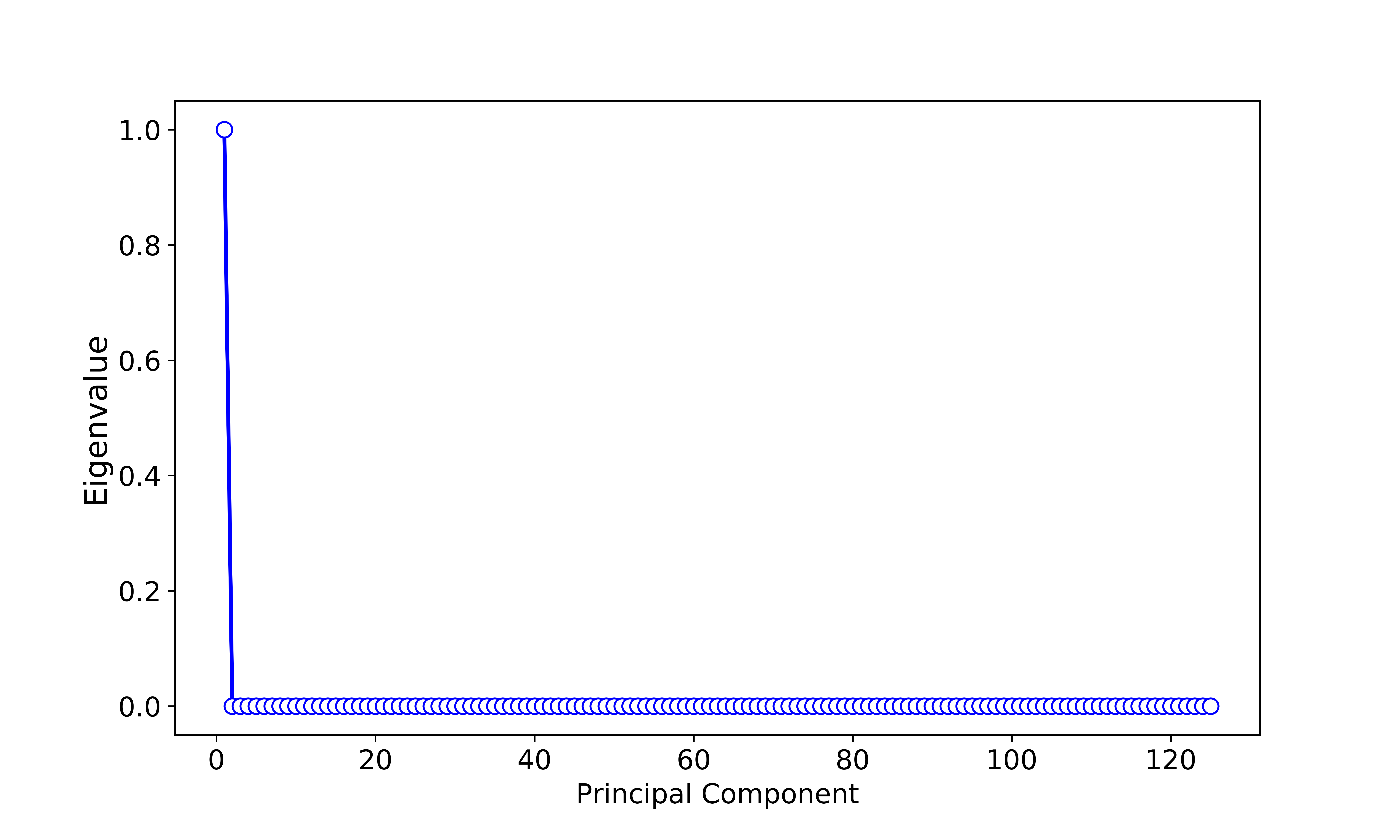}
\caption{Scree plot showing dominant contribution of PC1 and PC2.}
\label{fig:scree}
\end{figure}

To further validate component selection, a scree plot was generated (Figure~\ref{fig:scree}), showing a sharp drop in eigenvalue magnitude after PC2, confirming that additional components carry negligible variance and supporting the choice of a two-component representation.

\subsection{Visualization of Principal Components}
\begin{figure}[ht]
\centering
\includegraphics[width=0.9\linewidth]{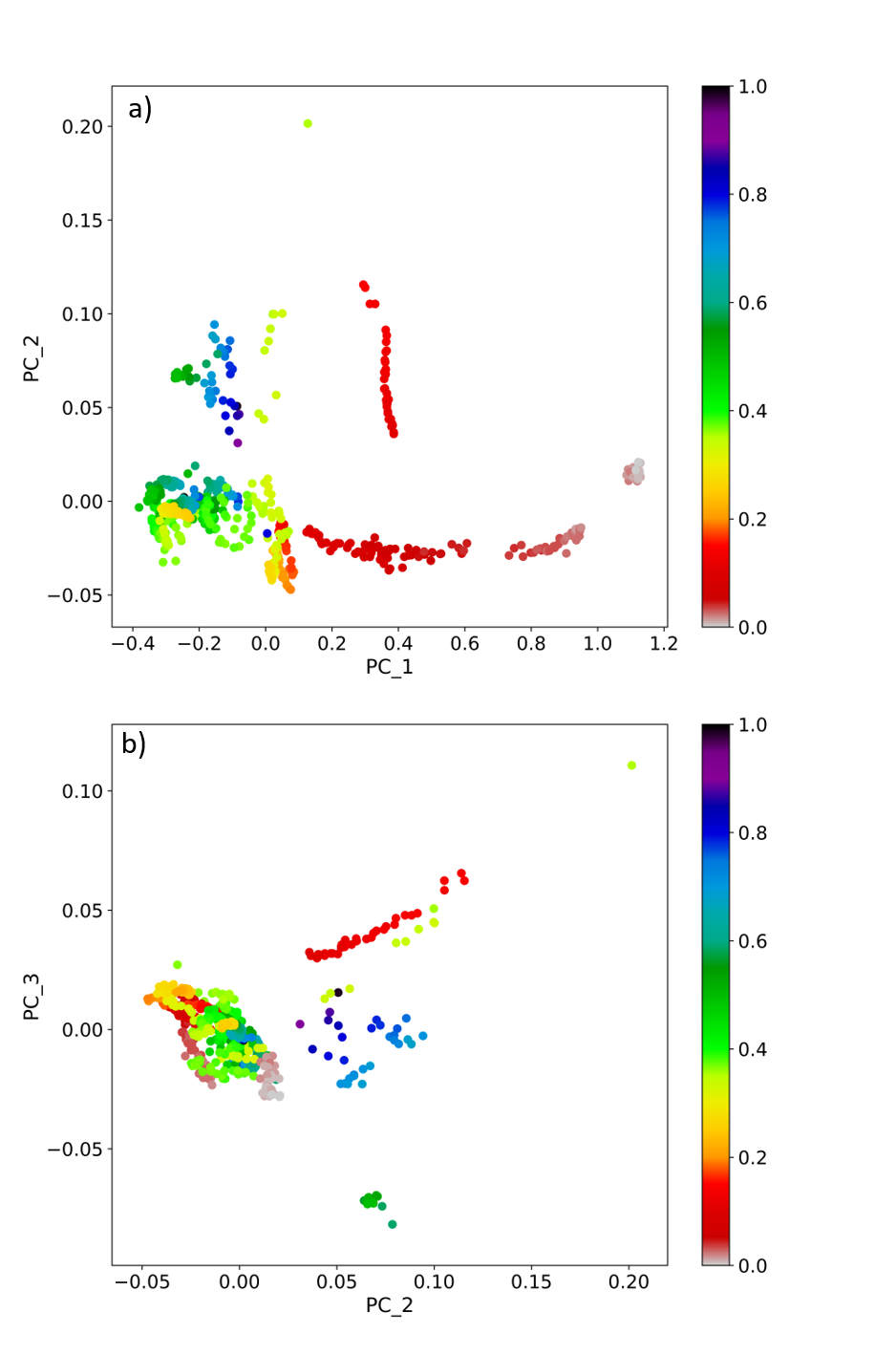}
\caption{(a) PC1 vs. PC2 and (b) PC2 vs. PC3 projections with soil moisture color mapping.}
\label{fig:pca_proj}
\end{figure}

Figure~\ref{fig:pca_proj} presents the projection of the dataset onto PC1 and PC2, with a color gradient corresponding to the normalized soil moisture values. The plot reveals a clear clustering trend, indicating that the two-component PCA transformation effectively separates data samples according to their soil moisture levels. This demonstrates that PCA can significantly simplify the representation of hyperspectral structure while retaining key predictive information.

Figure~\ref{fig:pca_proj} also shows PC2 vs. PC3 projection, where a noticeably stronger overlap between data points is observable, consistent with the very low explained variance of PC3 reported in Table~\ref{tab:pca}. Therefore, components beyond PC2 do not provide meaningful discriminatory power.

Finally, a baseline regression experiment was conducted using the \texttt{RandomForestRegressor}, configured with 100 trees. The model achieved a coefficient of determination \(R^2 = 0.947\), indicating that approximately 94.7\% of the variation in soil moisture can be explained by the hyperspectral features. This result highlights the strong predictive potential of optical spectral signatures for soil moisture estimation.

\section{Conclusion}
This study demonstrated that PCA-based dimensionality reduction can significantly enhance machine learning performance in hyperspectral optical imaging. By compressing 125 spectral bands into only two principal components while retaining over 99\% of the variance, PCA reduced redundancy, improved computational efficiency, and enabled clearer separation of spectral patterns relevant to the target variable. The results confirm that PCA is an effective strategy for simplifying high-dimensional hyperspectral datasets without compromising predictive accuracy, and it supports the development of more interpretable and scalable machine learning models for optical imaging applications. Future work should evaluate alternative feature extraction methods and test the framework across diverse imaging scenarios and sensor platforms.

\bibliographystyle{unsrt}

\end{document}